\documentclass[11pt,twoside]{article}   
\usepackage{latexsym} 
\usepackage{mathptm}
\usepackage{amsmath}
\usepackage{amssymb}
\usepackage{graphicx}
\usepackage{endnotes}

\def\beq{\begin{equation}}
\def\eeq{\end{equation}}
\def\beqn{\begin{eqnarray}}
\def\eeqn{\end{eqnarray}}

\begin{document}
 
\title{The Free Will Function\\
\large{Free will from the perspective of a particle physicist}}
\author{Sabine Hossenfelder \thanks{hossi@nordita.org}\\
{\footnotesize{\sl Nordita, Roslagstullsbacken 23, 106 91 Stockholm, Sweden}}}
\date{}
\maketitle
\vspace*{-.5cm}
\begin{abstract}
It is argued that it is possible to give operational meaning to free will and the process
of making a choice without employing metaphysics.
\end{abstract}

\section{Introduction}

The final session of the 2011 FQXi conference concluded with a brief survey. The question ``Is a `perfect predictor' of your choices possible?'' was answered with ``Yes'' by 17 out of 40 respondents. The follow-up question ``If there were, would it undermine human free will?'' was answered with ``Yes'' by 18 out of 38 respondents. 

It is striking how unrepresentative this result is for the general population who likes to hold on to the belief that personal choices are undetermined and unpredictable. In a cross-cultural study with participants from the United States, Hong Kong, India and Colombia, Sarkassian {\sl et al} \cite{Sark} found that more than two thirds of respondents (82\% USA, 85\% India, 65\% Hong Cong, 77\% Colombia) believe that our universe is indeterministic and human decisions are ``not completely caused by the past''(exact wording used in the study). 

There are two factors that likely contribute to Sarkassian {\sl et al}'s finding. One is that most people feel like they do have the freedom to make choices\endnote{The Diagnostic and Statistical Manual of Mental Disorders, Fourth Edition, ({\sc DSM-IV}) \cite{DSM}, describes Depersonalization Disorder as follows: ``The essential features of Depersonalization Disorder are persistent or recurrent episodes of depersonalization characterized by a feeling of detachment or estrangement from one's self. The individual may feel like an automaton or as if he or she is living in a dream or a movie. There may be a sensation of being an outside observer of one's mental processes, one's body, or parts of one's body.'' Thus, interestingly enough, not all of us share the feeling of being in charge of our actions.}. The other factor is that a rejection of free will has been claimed to be correlated with lowered moral responsibility \cite{Vohs}. 

However, neither of these supports of the common belief in free will is scientifically well founded. In fact, since rejecting free will may be a socially difficult position, those who do not believe in free will have an incentive to keep their conviction to themselves lest they cause social unrest or at least be considered odd. In light of this, it is even more interesting then that a selection of scientists puts forward an opinion significantly less in favor of the existence of free will than the general population.

The reason that such a large fraction of the above mentioned conference participants submit they believe to be predictable may be that it is difficult to embed a meaningful notion of ``free will'' into modern science. This paper is dedicated to the attempt to make sense of free will and to argue that one does not need to resort to metaphysics to make place for free will, though the argument will remain far from claiming that free will is real.

The next section explains the terminology and summarizes commonly made arguments. It is merely meant to clarify what we are talking about and does
not contain anything new. The third section offers, so I hope, an interesting new point of view.  

\section{Terminology}

Of course it all depends on what one means with ``free will,'' and since we don't want to waste time arguing about the meaning of words, let us start with clarifying the terminology.

There is a lot of recent research in neuroscience aimed at finding out whether we chose consciously or if we decide unconsciously before we become consciously aware of it \cite{Nature}. That is an interesting question, and one day science may force us to accept our brains not only do things we are not consciously aware of (which is the case for most of the brain's doings anyway) but that these unconscious brain ramblings are cause for decisions we cannot consciously change, or worse, that this is in fact the way our brain commonly operates, and conscious awareness is reduced to passive observation. However, for the question of free will it is, on the fundamental level, not particularly interesting exactly which part of the brain is involved in decision making or what you consider ``you''. Thus, we will in the following treat the human brain as a black box that operates on input and in reaction changes procedure, or produces output, but not further ask what parts of this decision are conscious and what are unconscious (which conveniently also avoids the necessity of explaining what consciousness is).
\subsection{Determinism}

With that preamble, a smallest common denominator of free will might then be put as follows:
\begin{itemize}
\item[1)] An agent in possession of free will is able to perform an action that was possible to predict by nobody but the agent itself.
\begin{itemize}
\item[a)]	In practice.
\item[b)]	In principle.
\end{itemize}
\end{itemize}
An ``agent'' here is to mean a system described by a collection of degrees of freedom, a subsystem of the universe, and an ``action'' is any change in its degrees of freedom ( i.e. an ``action'' can well be an entirely intrinsic, not necessarily externally observable, change). We will in the following refer to the values of all the agent's degrees of freedom as the ``state'' of the agent. 

The case 1a) in which a prediction is not possible in practice is too vague to be useful because it raises the question who can predict, when, by which methods, to which precision, in which time, does that prediction have to be faster than the agent's action itself, or at least not slower, did the planets have a free will before Kepler wrote down his laws, and other equally pointless distractions. We will thus in the following use the condition 1b), in which a prediction is not possible in principle. 

Our somewhat superficial notion of ``agent'' does not have a criterion of irreducibility in it, i.e. if we have a subsystem of the universe fulfilling 1b) and one not, the combination will fulfill 1b) too. Likewise, a group of agents with free will qualifies as another agent with free will. This problem could be fixed by more careful definitions that we however do not need  for our purpose, so we leave it at this.

The smallest common denominator 1) is a necessary condition for free will, but it does not seem sufficient for the question remains if the agent could have taken any other action. Even if nobody else is able to predict what the agent will do (in the simplest scenario that may be because no complete information about its degrees of freedom is available to anybody/anything able to process it) that does not mean the agent actually has a choice, it just means nobody knows what it will chose. To further tighten the case, we thus require

\begin{itemize}
\item[1')] An agent in possession of free will is able to perform an action that does not inevitably follow from all in principle available information at any time preceding the action.
\end{itemize}
Since the agent is a subset of the universe, this means an agent in possession of free will cannot exist in a universe with forward deterministic time evolution, and if such an agent existed, the time evolution would not be forward deterministic. 

``Forward deterministic'' means that the state of the universe at time $t$ determines the state of the universe for all $t' > t$, and thereby also the state of the agent. A backward deterministic time evolution similarly means that the state of the universe at time $t$ determines the state of the universe and with it that of the agent for all $t'<t$. 

In classical physics one usually has a time evolution that is forward and backward deterministic (or just deterministic), in which case it uniquely maps states at any one time into states at any other time and is also time reversible -- this is the universe of Laplace's demon who is able to predict all the future if just given sufficient initial data at any one time. Quantum mechanics in the standard interpretation (the measurement collapses the wavefunction) is neither forward nor backward deterministic: One can only predict the probability of the outcome of a measurement, but not the outcome itself (not forward deterministic) and once one has made a measurement one cannot reconstruct the state previous to measurement (not backward deterministic). If the evaporation of black holes is indeed completely thermal, then it is a forward deterministic process (it is not reversible and in particular not unitary).

Since an evolution that is not forward deterministic is in principle unpredictable, it it follows 1') $\Rightarrow$ 1b), though the reverse it not necessarily true: It may not be possible to predict what happens to a collection of particles we cannot interact with, yet that does not mean their time evolution is indeterministic. 1') is thus a stronger statement than 1b). 

Note that the time evolution of the agent, as a not necessarily closed subsystem of the universe, will in general be non-deterministic if one takes as initial state only the state of the agent at an earlier time, because then its interactions with the environment make an unknown contribution. (I can't predict what you'll be having for lunch tomorrow if I don't know what you'll read on the menu.)

\subsection{Compatibilism}

The rationale of 1') is that if time evolution was forward deterministic then the agent's actions at a time $t_0$ would have been decided already at any earlier time $t < t_0$ (if necessary by all the states in the universe) and its future evolution (or call it behavior) would be entirely unchangeable and had been fixed already at the Big Bang (or earlier, depending on your favorite cosmological model). It does not seem to make much sense to speak of free will in this case, though ``the majority of philosophers today maintain one or another form of compatibilism'' \cite{Sark} that is the belief that free will is indeed compatible with determinism. And who are we to argue with them about the use of a word? It seems then though that we've lost the majority of philosophers by tightening our case from 1) to 1'), but if you are among the compatibilists, stay with us for a while, for the argument will become more compatible later on. 

One commonly found rationale for compatibilism stems from the idea that while the fundamental laws of physics may dictate time evolution on the microscopic level, our agent's ``free will'' is an emergent property which may obey emergent laws on the macroscopic level. To our best present knowledge any macroscopic law would in principle strictly follow from the microscopic ones (though they might in practice be difficult to derive) so they would be equally deterministic and not open any room for free will. But even if one does not believe the macroscopic laws follow from the microscopic ones and more really is different (which remains to be shown, but see \cite{different}), then one just removes the relation to fundamental law, but 1') remains a statement about the evolution of the macroscopic states of the agent, which then better not be deterministic. Thus, calling upon emergence does not help for then we will just ask if the emergent laws are deterministic or what else. Neither does it help to attach adjectives like ``autonomous'' or ``complex'' or ``chaotic,'' since none of that renders a deterministic time evolution non-deterministic.

One might just insist to call something ``free will'' that fulfills 1b) but not 1') and say that one doesn't care if a choice was predetermined so long as nobody else can actually predict it. One might even be satisfied with 1a). Fighting about words is moot and it is questionable we have any choice with our opinions anyway, so in the following 1') should be considered just a definition. To differentiate 1) and 1'), the philosopher Edward Fredkin suggested that the 1b) type of free will that exist in a deterministic universe and does not fulfill 1') be named ``pseudo free will'' \cite{Fredkin} and David Albert called it, somewhat less judgmental, the ``inviolable private will'' \cite{Albert}. 

At this point one may also note that if time evolution is reversible, one might equally well say that an agent's actions in the past were determined by the future, just that, with the arrow of time pointing towards wrinkles, that's not how we deal with the world around us. It is simply impractical since we do not have memories of the future and we are able only to extrapolate the past forward, but not the future backwards. That we are not able to do so does however not mean the laws of nature forbid it.

Now that we have argued free will needs an indeterministic time evolution, we unfortunately have to note that a random element in the time evolution does not concur with our cherished idea of free will either. According to best present knowledge, quantum mechanics tells us the time evolution of a block of radioactive material is in principle not predictable and in fact not deterministic. All we can do is to find out the half-life and give a statistic evolution law for the decay rate, yet we do not take this as evidence that radioactive material has a free will. If decisions would be
random, the agent would not be in charge of them in any meaningful way. So we add another requirement:

\begin{itemize}
\item[2)] The actions of an agent in possession of free will cannot be consequences of fundamentally random processes.
\end{itemize}

By ``random'' we mean random in the mathematical sense, i.e. given by a random variable which has no distinct value but which can take on a set of values with assigned probabilities. 

Having come so far, it is clear why almost half of the respondents, of the {\sc FQX}i survey did not believe they are able to freely make decisions. Since we do at present not have one single candidate theory for the fundamental laws of nature that is neither deterministic nor random, it is difficult to maintain a belief in free will. It does not help to combine deterministic with stochastic evolution. Such a combination might make one or the other
aspect less apparent, but still does not offer us a way to make sense about an agent freely making decisions in accordance with 1') and 2).  

\subsubsection*{Brief Excursion: Free Will in Quantum Mechanics}

Before we continue, a few words on the ``Free Will Theorem'' \cite{fwt}. In brief, the Free Will Theorem is concerned with the notion of free will as commonly used in quantum mechanics, typically the freedom of the experimenter to decide how to set the detector. The Free Will Theorem states that if the experimenters have free will in that sense, then so must elementary particles. A logically equivalent statement is that if elementary particles do not have free will then experimenters have no free will either. Thus, contrary to what the name ``Free Will Theorem'' lets one hope, it actually does not deliver free will by merit of going from classical to quantum mechanics. 

In fact, the free will of the experimenter is an assumption to Bell's theorem, the very theorem that has been used to falsify deterministic locally realist hidden variable models, thereby establishing non-deterministic quantum mechanics (though loopholes remain []). Thus, while the non-deterministic evolution of quantum mechanics breaks with Laplace's demon who is able to predict the motion of all and everything if just given initial conditions and makes room for free will, an argument for free will to find its place in the non-determinism of quantum mechanics is circular: It necessitates free will to make room for free will. 

In other words, just going from classical to quantum mechanics does not make us any wiser.

\subsection{Libertism}

We have seen that if we force ourselves to make mathematically concrete the intuitive but fuzzy concept of free will, then that concept is missing in all laws of nature that we know and use today. If one is not content with the restrained 1) version of free will that is compatible with determinism, then one faces the question what reality would have to look like in order for free will to be more than an illusion. 

The cheapest way out is to hypothesize that there are things science can't describe, that reality can't be cast in mathematics, that there is some metaphysical something that can't follow from any law of nature, and so on. These are all desperate attempts, known as ``libertism,'' for which there is not only absolutely no scientific evidence but, worse, they are non-scientific by construction. However, in the following we argue is possible to hold on to a notion of free will without retreating to metaphysics and that there is a version of libertism compatible with modern science.

Again, one could mention that the words ``determinism,'' ``compatibilism'' and ``liberalism'' are used here differently from how they have been used elsewhere. That is likely true, but beside the point as we've just used them to name concepts. Stripped of  -isms, the statement is the following: It is possible to have a time evolution in which an agent's future actions are not completely determined, even in principle, by all information available in the present and in the past, and its actions are not random either in the sense that the value of  variables initiating a change (action) are distinct.

\section{The Free Will Function}

\begin{figure}[ht]
\includegraphics[width=8.0cm]{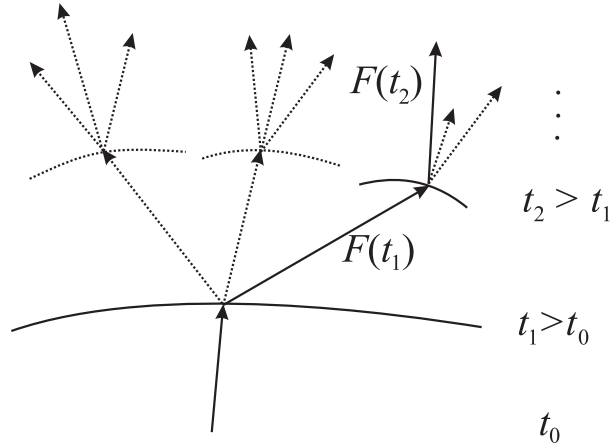}
\caption{{\small One branch of the history that is not fully specified by $H(t)$ (dotted and solid) is selected by the free will function $F(t_i)$ (solid).}}
\label{4}
\end{figure}

We have argued above that in order for free will to be more than an illusion, we need a time evolution that is neither deterministic nor random. 
We can take this to mean the following: There is a time evolution $H(t)$ that is not forward deterministic in the sense that given the agent's state at some time $t_0$, $H(t)$ allows for a set of states at time $t_1 > t_0$. For simplicity, let us assume that the evolution is reversible and deterministic except for a series of moments, $t_i$, $i \in N$, in which the agent ``makes a decision'' and the set of possible states branches (see figure \ref{4}) into different options that are only probabilistically known. Let us also assume that each decision comes down to choosing one from ten alternatives described by the digits 0 to 9. 

What we need in order for this evolution to not be random is a function $F(t_i)$ that we can call the ``free will function'' that at any time $t_i$ returns a specific choice, i.e. a digit, and by that selects a uniquely specified path.  The evolution $H(t)$ then becomes deterministic if one selects the path with help of $F$ in every branching point. The free will function $F(t_i)$ assigned to the agent one could loosely interpret as the agent's character. One should not take this interpretation too seriously since human character has changeable components that evolve over time, and the free will function
would only be some aspect of it that might vary in relevance.  

The function $F$ should not be forward deterministic itself, otherwise we would be back in the block universe with Laplace's demon. Neither should it be a random process.  Functions of this type are not difficult to find. One just needs an (in principle) uniquely specified function that cannot be evolved forward for example because it makes discrete jumps or because it is not forward deterministic. That function must in addition have the property that it cannot be constructed just by collecting a (possibly infinite) set of values to the past.
 
Here is an example: Consider an algorithm that computes some transcendental number, $\tau$, unknown to you. Denote with $\tau_n$ the $n$-th digit of the number after the decimal point. This creates an infinitely long string of digits. Let $t_N$ be a time very far to the future, and let $F$ be the function that returns $\tau_{N-i}$ for the choice the agent makes at time $t_i$. 

This has the following consequence: The time evolution of the agent's state is now no longer random. It is {\it determined} by $F$, but not (forward) {\it deterministic}: No matter how long you record the agent's choices, you will never be able to predict, not even in principle, what the next choice will be (1b).  If you constrain yourself to recording his choices for a finite amount of time you wouldn't be able to predict his next move even if you knew $\tau$. You could go and ask Laplace's demon to tell you all the degrees of freedom of the whole universe at any one time and still you would not be able to figure out what the agent will be doing next (1'). But if you only knew the agent's free will function, there would be nothing ambiguous about his actions (2). In some sense you could say, the agent ``is'' his free will function. Any time our agent faces a decision, it calls upon its free will function and asks for the next number, which corresponds to it ``making a choice.''

The status of the free will function here is like that of a fundamental law of nature. It is not caused by anything else, it does not follow from anything, it just ``is.'' 

Another possibility is to complement a forward deterministic evolution $H(t)$ with a backward deterministic evolution $F(t)$, which would have the effect that to be able to find out what happened at any time $t_0$, one would need an initial state at a time $t<t_0$ (for the forward evolution) and a final state at a time $t>t_0$ (for the backward evolution). Since we only have access to initial states in the past, we would never be able to know what happens. That example is in some sense less satisfactory since it comes close to saying if one knows what will happen then one can predict what will happen, which is tautologically true but not a very deep insight.

Of course these examples are arbitrarily constructed and are certainly not meant to describe actual reality. Their purpose is merely to show that it is possible to have a mathematical description of reality that does allow for free will to exist and give operational sense to the act of making a decision in a world that is determined but not deterministic. This admittedly opens more questions than it answers: How can one embed such a free will function into the currently known laws of physics? How can we find out if we have such a free will? If we have such a free will, why do we have it and who or what else can have it? 

There is also an interesting lesson to be learned from this. All examples that allow for free will have in common that the free will function cannot be a solution (at least piecewise) to a differential equation for if it was it could be evolved forward by use of this equation. If one believes that Nature is described by mathematical law there is no specific reason why this law should come in form of a differential equation -- except that this is what our brains, trained by evolution and education, most naturally extract from observations, and in fact what all modern theories make use of: Given some initial configuration we ask what will happen to it in the future.

But one can in principle specify a function (that we may imagine describes our universe) by a set of axioms, none of which is that it be a solution to a differential equation. If one had these axioms rather than the differential equation, then instead of finding a solution and suitable initial data, one would need to specify a location in the space over which the function is defined. A fundamental theory that is not formulated by use of differential equations may not need initial conditions, thus opening a new direction to address the question why our universe started the way it did. The answer may be that theories using differential equations that work well locally reach their limits when we aim to describe the history of the whole universe, and interestingly enough the existence of free will, should it exist, might be a hint of this.
\section{Conclusion}

We have argued that it is possible to make scientific sense of free will and have suggested an operational meaning for ``making a choice.'' An agent can make a choice that does not follow from any information available in the past, by reading out the value of a ``free will function'' that has to fulfill the only requirement of not being forward deterministic. 

\section*{Acknowledgements}

I thank Joy Christian, Lee Smolin, Stefan Scherer and Scott Aaronson for valuable feedback, and all participants of the
FQXi 2011 conference for making this meeting so enjoyable.

\theendnotes

\end{document}